\documentstyle[12pt,a4,cite,axodraw,epsfig,amssymb,amsmath]{article}

\newcommand{\lsi}{\raise0.3ex\hbox{$<$\kern-0.75em\raise-1.1ex\hbox{$\sim$}}}
\newcommand{\gsi}{\raise0.3ex\hbox{$>$\kern-0.75em\raise-1.1ex\hbox{$\sim$}}}
\newcommand{\lsim}{\mathop{\lsi}}
\newcommand{\gsim}{\mathop{\gsi}}
\newcommand{\tb}{\tan\!\beta}
\newcommand\gd\delta
\newcommand{\de}{\partial}
\makeatletter \@addtoreset{equation}{section} \makeatother
\renewcommand{\theequation}{\arabic{section}.\arabic{equation}}
\begin{document}

\setlength{\baselineskip}{0.6cm}
\begin{titlepage}
\begin{flushright}
HD-THEP-00-04\\
hep-ph/0002050
\end{flushright}
\begin{centering}
\vfill

{\bf Do Stops Slow Down Electroweak Bubble Walls?}
\vspace{1cm}

P. John\footnote{P.John@thphys.uni-heidelberg.de} and
M.G. Schmidt\footnote{M.G.Schmidt@thphys.uni-heidelberg.de} \\

\vspace{1cm} {\em 
Institut f\"ur Theoretische Physik, 
Philosophenweg 16, 
D-69120 Heidelberg, Germany
}

\vspace{2cm}

{\bf Abstract}

\end{centering}
\vspace{0.3cm}\noindent
We compute the wall velocity in the MSSM. We therefore generalize the
SM equations of motion for bubble walls moving through a hot plasma at
the electroweak phase transition and calculate the friction terms
which describe the viscosity of the plasma. We give the general
expressions and apply them to a simple model where stops, tops and W
bosons contribute to the friction. In a wide range of parameters
including those which fulfil the requirements of baryogenesis we find
a wall velocity of order $v_w\approx (5-10)\cdot 10^{-2}$ much below the
SM value.
\vfill 
\vfill
\noindent

\end{titlepage}

\section{Introduction}
The generation of the baryon asymmetry of the Universe in a first
order phase transition (PT) of the electroweak theory is attractive
because all of its ingredients could be tested in high energy
experiments in the near future. However, it is for sure now that there
is no strong first order PT in the Standard Model (SM) with a Higgs
mass above the experimental lower bound, indeed there is no PT at all
\cite{lattice98}. The SM is very successful, but there is common
agreement that it has to be extended to a more general theory. That
this theory will contain supersymmetry is still controversial but in
lack of alternatives this is a very useful hypothesis. In the MSSM and
also in an extension with an additional singlet field (NMSSM)
\cite{frog,HuberSchmidt,HS2} it is possible to obtain a strong first
order PT without violating the experimental Higgs mass bounds.  In the
MSSM this is related to a scalar partner of the right handed top which
is very light in the symmetric phase \cite{John2,litestop,Cline12}.
It increases the cubic term in the effective 1-loop scalar
potential. In the NMSSM a $SH_1H2$ term arises already at tree level
and acts effectively as a $\phi^3$-term if the vacuum expectation
value (vev) ${<\!S\!>}$ is comparable to the Higgs vevs
\cite{PietroniNMSSM}. These types of models have also more freedom to
realize CP-violating effects.  It is well known that supersymmetric
models allow for spontaneous CP-violation at $T=0$. The parameter
space is strongly restricted by experimental bounds on the electric
dipole moment of the neutron (for a discussion and references see
e.g. \cite{WagnerHiggse}).  Moreover to produce a baryon asymmetry at
the electroweak scale CP-violation within the bubble wall is needed.
Therefore it has been proposed that a temperature induced transitional
CP-violation might occur
\cite{CPRiotto,JapanerSpezialfall,Mikko3dCP}. In \cite{John4,JohnPhD}
it was shown that in the MSSM spontaneous CP-violation does not occur
throughout. Even with maximal explicit CP violating phases the
variation of the corresponding phase in the Higgs system is strongly
suppressed.

The baryon asymmetry arises in a two step process: first the expanding
wall sweeps through the hot plasma separating Higgs phase and
symmetric phase with a CP-violating spatially varying Higgs vev.  It
generates an asymmetry between left handed quarks and their
antiparticles diffusing in front of the bubble starting from an
asymmetry between stops, tops, charginos, neutralinos and their
antiparticles. In a second step this asymmetry in the hot plasma in
front of the bubble wall is transformed into a baryon asymmetry
through (hot) unsuppressed sphaleron transitions. If the PT is
strongly first order this asymmetry is not destroyed by the Higgs
phase (weak) sphaleron when finally the equilibrium phase takes over
. The rise of the generated baryon asymmetry depends on the spatial
variation of the CP-violating phase in the wall. Moreover, in the MSSM
the variation $\Delta\beta\sim d\beta/dx$ of $\tb$ is an
important ingredient for the determination of the baryon asymmetry
$\eta$. Due to its smallness the observed baryon asymmetry may be
reached only in a small, eventually tuned range \cite{dfl} of
parameters and therefore also a better knowledge of the wall velocity
is needed, since there is a strong dependency on it. For instance in
\cite{CQViljaW} it was found in semiclassical calculations that with a
given Higgs field profile $h^2(x)$
\begin{equation}
\eta\sim\int_{-\infty}^{\infty}dx\frac{h^2(x)}{v_wT_c^2}\frac{d\beta(x)}{dx}.
\end{equation}
Thus the asymmetry increases with decreasing wall velocity. Of course
$v_w\to 0$ is not possible, in this case there would be no
out-of-equilibrium region.  Quite naively one might expect that more
particles would imply more interactions and in consequence a higher
viscosity of the plasma. For this reason it usually was assumed
(e.g.\cite{brhlik}) that the velocity of bubble walls of
supersymmetric phase transitions is smaller than the velocity in the
SM. But that has to be shown in a detailed calculation.

To go beyond speculations we want to reconsider this question and the
calculation of the wall velocity taking into account also
supersymmetric particles. The proceeding is similar to that of
\cite{MP12}. Nevertheless, in case of two (Higgs-) scalars and
arbitrarily many interacting particle species (fermions {\em and}
bosons), the calculation and the results differ considerably from
\cite{MP12}. Therefore, we will explicitely outline the main steps
of the calculation (see also \cite{JohnPhD}) to demonstrate the
differences. On the other hand, at some points which can be found in
\cite{MP12} we keep short in the representation.
\section{Bubble Wall Equation of Motion}
Energy conservation leads to the equations of motion of an electroweak
bubble wall interacting with a hot plasma of particles \cite{Linde,LMT,MP12}:
\begin{equation}
\square h +V'(h)+\sum_i \frac{\de m_i^2}{\de h}\int\frac{d^3k}{(2\pi)^3 2E}f_i(k,x)=0,
\label{walleqmo1}
\end{equation}
where $f_i=f_{0,i}+\delta f_i$ is the distribution function for a
particle species in the heat bath. We have to sum over all particle
species $i$.  The distribution function is divided up into equilibrium
part $f_{0,i}$ and out-of-equilibrium part $\delta f_i$. For two scalars
(or more, analogously) we obtain
\begin{eqnarray}
\square h_1 + \frac{\de V_T(h_1,h_2)}{\de h_1} + \sum_i\frac{\de m_i^2}{\de h_1}\int\frac{d^3p}{(2\pi)^32E_i}\delta f_i(p,x) &=& 0,\label{walleqmo2}\\
\square h_2 + \frac{\de V_T(h_1,h_2)}{\de h_2} + \sum_j\frac{\de m_j^2}{\de h_2}\int\frac{d^3p}{(2\pi)^32E_j}\delta f_j(p,x) &=& 0.\label{walleqmo3}
\end{eqnarray}
The equilibrium part has been absorbed into the equilibrium
temperature dependent effective potential $V_T(h_1,h_2)$. In
equilibrium we would obtain the free equations for critical bubbles
respectively domain walls for large radii.

In the following we will restrict ourselves to late times leading to a
stationarily moving domain wall where the friction stops the bubble
wall acceleration. This is the long period of bubble expansion where
baryogenesis takes place. The influence of different friction or
viscosity terms in Standard type models was investigated in various
papers \cite{LMT,Linde,Pantano}.  We assume not too large velocities
and check the self consistency of this assumption in the MSSM.
\section{Fluid Equations}
\label{sec:fleqs}
Now we want to derive the deviations $\delta f_i$ from the equilibrium
population densities originating from a moving wall. We will therefore
discuss the Boltzmann equation in the fluid frame, the ''fluid
equations'':
\begin{equation}
d_tf_i\equiv \de_tf_i + \dot x \frac{\de}{\de x} f_i + \dot p_x \frac{\de}{\de_{p_x}} f_i = -C[f_i],
\end{equation}
with the population density $f_i$ and energy $E=\sqrt{p_x^2 + m^2(x)}$.
$C[f_i]$ represents the scattering integral and will be discussed in
section~\ref{sec:graphs}. The classical (WKB) approximation is valid
for 
\begin{eqnarray}
p&\gg& \frac{1}{L_w} \quad\mbox{(``thick wall'')}.
\label{WKB}
\end{eqnarray}
For particles with $E,p \gsim gT$ this should be fulfilled.  Thus,
here infrared particles are supposed not to contribute to the friction
\cite{MP12,LMT,Linde}. This is a crude approximation.  A further
understanding of the infrared particle contribution is therefore
needed which goes beyond the aim of this
paper\footnote{Ref.~\cite{Moorewall2000} revisits the calculation of
the friction in the SM. The prediction is that hard thermal loop
effects play an important role in the damping of the gauge fields in
the hot phase and that such coherent gauge field contributions are
very effective for generating a friction. }.
In the MSSM the wall thickness $L_w$ is of order $15/T$--$40/T$, as
found in \cite{SecoNum,Cline12,JohnPhD}, and $L_w\gg 1/T$ is
fulfilled. With \cite{MP12} we denote those particles which couple
very weakly to the Higgs as ``light particles''. Particles coupling
strongly to the Higgs are heavy in the Higgs phase and therefore
called ``heavy''. However, ``superheavy'' particles as the ``left
handed'' stops do not appear in our calculation besides their remnants
in the effective potential.  We treat as ``heavy'' particles only top
quarks, (right handed) stops, and W bosons. Besides for the physical
stop mass, we neglect the $U(1)$ and treat only the $SU(2)$ with its
coupling $g$. The Higgses are left out.

We assume now that the interaction between wall and particle plasma is
the origin of small perturbations from equilibrium. We will treat
perturbations in the temperature $\delta T$, velocity $\gd v$ and
chemical potential $\delta\mu$ and linearize the resulting fluid
equations. Then the full population density $f_i$ of a particle
species $i$ in the fluid frame is given by
\begin{equation}
f_i= \frac{1}{\exp\left\{\frac{(E+\gd_i)}{T} \right\}\pm 1}\label{Eplusd},
\end{equation}
where we have generally space dependent perturbations $\gd_i$ from
equilibrium. In principle one must include perturbations from the
global value for each particle species. A simplification is to treat
all the ``light'' particle species as one common background fluid.
This background fluid obtains common perturbations $\gd v_{bg}$ in the
velocity and $\gd T_{bg}$ in the temperature. This leads us to
\begin{equation}
\gd_i= -\left[ \delta\mu_i + \frac{E}{T}(\gd T_i + \gd T_{bg}) + p_x(\delta v_i+\delta v_{bg})\right]
\label{gd}
\end{equation}
for the ``heavy'' particles. The spatial profiles of all these
perturbations depend on the microscopic physics.  We treat particles
and antiparticles as one species neglecting CP violation which is a
minor effect on the friction. For the calculation of the baryon
asymmetry this is the important effect and would be involved in
eq.~(\ref{gd}) by a perturbation $\gd E_\pm$ in the energy dividing
particles and antiparticles \cite{baryogen,HS2}.

We now expand $d_tf$ to linear order in the perturbations. The
Boltzmann equation can then be written as
\begin{eqnarray}
&&(-f_0^\prime)\left( \frac{p_x}{E}[\de_x\delta\mu  +\frac{E}{T}\de_x(\gd T + \gd T_{bg}) +p_x\de_x(\delta v +\delta v_{bg})]+\de_t\ \delta\mu \right. \nonumber\\
&&  \left.+  \frac{E}{T}\de_t(\gd T + \gd T_{bg}) +p_x\de_t(\delta v+\delta v_{bg})\right) +TC(\delta\mu,\gd T,\delta v) = (-f_0^\prime)\frac{\de_t(m^2)}{2E}.\nonumber\\
\label{B1}
\end{eqnarray}
The term on the right hand side of (\ref{B1}) drives the population
density away from equilibrium.

The collision term depends on all perturbations. But since the
perturbations are Lagrangian multipliers for particle number, energy,
and momentum, we can determine the parameters by the appropriate
integration choice
\begin{equation}
\int \frac{d^3p}{(2\pi)^3},\quad\int E\frac{d^3p}{(2\pi)^3},\quad\mbox{and}\quad\int p_x\frac{d^3p}{(2\pi)^3}.\label{B2}
 \end{equation} 

The resulting three equations coming from the Boltzmann equation, the
``fluid equations'', are coupled through the collision term
$C[\delta\mu,\delta T,\gd v]$. Performing the integrals leads to the
general pattern
\begin{eqnarray}
\int\frac{d^3p}{(2\pi)^3T^2} C[f] &=& \delta\mu\Gamma_{\mu_1} + \gd T \Gamma_{T_1},\nonumber\\
\int\frac{d^3p}{(2\pi)^3T^3}E C[f] &=& \delta\mu\Gamma_{\mu_2} + \gd T \Gamma_{T_2},\nonumber\\
\int\frac{d^3p}{(2\pi)^3T^3}p_x C[f] &=& \gd vT\Gamma_v,\label{gammadef}
\end{eqnarray}
where the rates $\Gamma$ are of the form $\Gamma\sim
\alpha^2\ln(1/\alpha)T$, and $\alpha=g^2/(4\pi)$ is the gauge coupling.

The expressions (\ref{gammadef}) have to be evaluated graph
by graph through the out of equilibrium interactions of each particle
species. In sec.~\ref{sec:graphs} we will calculate the leading
contributions. 

For a stationary wall we can use $\de_t\delta_i\to v_w\delta'_i,$ and
$ \de_z\delta_i\to \delta'_i$, where the prime denotes the derivative
with respect to $z=x-v_wt$ (Galilei transformation for small $v_w$).

In the following the equations are again similar to those in \cite{MP12} but there are
important additional terms. We can solve the fluid equations to
eliminate the background perturbations $\gd v_{bg}^\prime$ and $\gd
T_{bg}^\prime$:
\begin{eqnarray}
\gd T_{bg}^\prime &=& \frac{-v_w(A+B)+ C + D}{\bar{c}_4(\frac{1}{3}-v_w^2)} \label{Tbg},\\
\gd v_{bg}^\prime &=& \frac{A+B - 3v_w(C + D)}{T\bar{c}_4(\frac{1}{3}-v_w^2)},\label{vbg}
\end{eqnarray}
where we have used the notation
\begin{eqnarray}
A &=& \sum_{\mbox{fermions}\ f} N_f(\delta\mu_f\Gamma_{\mu 2f} + \gd T_f\Gamma_{T2f}),\nonumber\\
B &=& \sum_{\mbox{bosons}\ b} N_b(\delta\mu_b \Gamma_{\mu 2b}+ \gd T_b\Gamma_{T2b}),\nonumber\\
C &=& \sum_{\mbox{fermions}\ f} N_f\gd v_fT\Gamma_{vf},\nonumber\\
D &=& \sum_{bosons\ b}N_b\gd v_bT\Gamma_{vb}.
\end{eqnarray}
In contrast to the computations in \cite{MP12} these terms may include
arbitrarily many fermions and bosons, respectively.

We can see that for walls moving with the velocity of sound
$v_w=v_s=1/\sqrt{3}\approx 0.577$, the approximation obviously breaks
down. This demonstrates the limit of the expansion in linear
perturbations.

For each heavy particle species in the plasma we have three fluid
equations resulting from the combination of eqs.~(\ref{B1}),
(\ref{B2}) and (\ref{gammadef}).  The final form of the fluid
equations can be written in a matrix notation:
\begin{equation}
{\mathbb{A}}\delta'+\Gamma\delta=F,
\label{fluidfinal}
\end{equation}
where 
\begin{equation}
\Gamma=\Gamma_0+\frac{1}{\bar{c}_4}\mathbb{M}.
\end{equation}
The matrices $\mathbb{A}$, $\Gamma$, $\Gamma_0$, and $\mathbb{M}$ are
given below. The number $\bar{c}_4$ is the heat capacity of the plasma
$\bar{c}_4=78 c_{4-}+37c_{4+}$ including light quarks, leptons, and
sleptons in the plasma. The number changes when we include further
light particles.  The perturbations are combined in a vector $\delta$,
the driving terms are combined in the vector $F$. The driving term
containing $(m^2)'$ can be split up into different contributions 
\begin{equation}
(m^2)' = \frac{\de m^2}{\de h_1}h_1' + \frac{\de m^2}{\de h_2}h_2'.
\label{msquared}
\end{equation} 
The vector $\delta$ and the matrices for $k$ particle
species (index $x$ denotes $+$ or $-$, for fermions and bosons,
respectively, for the $i$th particle):
\begin{eqnarray}
\gd=\begin{bmatrix}
\gd\mu_1 & \gd T_1 & T\gd v_1 & \ldots &\gd \mu_k&\gd T_k & T\gd v_k\end{bmatrix}^T
,\\
 F=\frac{v_w}{2T}\begin{bmatrix}
c_{1x}(m_1^2)'&c_{2x}(m_1^2)'& 0 &\ldots &&c_{1x}(m_k^2)'&c_{2x}(m_k^2)'&0\end{bmatrix}^T.
\end{eqnarray}
The matrices $\mathbb{A}$ and $\Gamma$ are of block diagonal form:
\begin{eqnarray}
\mathbb{A}=\begin{bmatrix}
{\mathbb{A}}_1 &  & \\
             &  \ddots & \\
             &  &  {\mathbb{A}}_k
\end{bmatrix},&\mbox{where}& {\mathbb{A}}_i=\begin{bmatrix} v_wc_{2i} & v_wc_{3i} &\frac{c_{3i}}{3}\\
                                                              v_wc_{3i} & v_wc_{4i} &\frac{c_{4i}}{3}\\
                                                              \frac{c_{3i}}{3} & \frac{c_{4i}}{3} & \frac{v_wc_{4i}}{3}
\end{bmatrix},\\
&&\nonumber\\
\Gamma_0=\begin{bmatrix} \Gamma_1 & &  \\
                         & \ddots & \\
                         & &  \Gamma_k
\end{bmatrix},&\mbox{where}& \Gamma_i=\begin{bmatrix}
                                        \Gamma_{\mu 1,i} &\Gamma_{\mu2,i} & 0\\
                                        \Gamma_{T1,i}    &\Gamma_{T2,i}   & 0\\
                                        0 & 0 & \Gamma_{v,i}
                                \end{bmatrix}.\label{gamma0}
\end{eqnarray}
Moreover, $\mathbb{M}$ is a square matrix of the form 
\begin{equation}
\mathbb{M}=\begin{bmatrix}
M_{1,1} &  \cdots & M_{1,k}\\
\vdots & \ddots & \vdots& \\
M_{k,1}  & \cdots & M_{k,k}
\end{bmatrix},
\end{equation}
\begin{equation}
M_{i,j} = N_i\begin{bmatrix}
                c_{3i}\Gamma_{\mu2,j} & c_{3i}\Gamma_{T2,j} & c_{3i}\Gamma_{v,j}\\
                c_{4i}\Gamma_{\mu2,j} & c_{4i}\Gamma_{T2,j} & c_{4i}\Gamma_{v,j} \\
                c_{4i}\Gamma_{\mu2,j} & c_{4i}\Gamma_{T2,j} & c_{4i}\Gamma_{v,j} 
                \end{bmatrix},\quad i,j=1\ldots k,
\label{Mij}
\end{equation}
where $i$ in $c_{3i}$, $c_{4i}$ denotes fermionic or bosonic
contributions $c_{3\pm}$, $c_{4\pm}$ of the $i$th and $j$th particle
species, respectively. They are defined for fermions(+) and bosons(-)
through
\begin{equation}
c_{n\pm}= \int \frac{E^{n-2}}{T^{n+1}}f_0^\prime(\pm)\frac{d^3p}{(2\pi)^3}.\label{definitionc}
\end{equation}
\section{Higgs Equation of Motion and Friction}
\label{sec:fricandeqmo}
With the definition (\ref{definitionc}) and taking into account
(righthanded) stop-, top- and $W$ particles the equations of motion
can be approximated in the fluid picture as
\begin{eqnarray}
0 &=& -h_1'' + \frac{\de V_T}{\de h_1}(h_1,h_2) +\frac{N_W}{2}T\frac{dm^2_W}{dh_1}(c_{1-}\delta\mu_b+c_{2-}\gd T_W+c_{2-}T\gd v_W),\nonumber\\
0 &=& - h_2'' +\frac{\de V_T}{\de h_2}(h_1,h_2) +\frac{N_W}{2}T\frac{dm^2_W}{dh_2}(c_{1-}\delta\mu_W+c_{2-}\gd T_W  +c_{2-}T\gd v_W      )\nonumber\\
 &+& \frac{N_t}{2}T\frac{dm^2_t}{dh_2}(c_{1+}\delta\mu_t+c_{2+}\gd T_t +c_{2+}T\gd v_t  )\nonumber\\
 &+&  \frac{N_{{\tilde t}_1}}{2}T\frac{dm^2_{{\tilde t}_1}}{dh_2}(c_{1-}\delta\mu_{\tilde{t}}+c_{2-}\gd T_{\tilde{t}}  +c_{2-}T\gd v_{\tilde{t}}).
\end{eqnarray}
This can formally be rewritten as
\begin{eqnarray}
-h_1''+\frac{\de V_T}{\de h_1}(h_1,h_2,T) + \frac{T}{2}h_1 G_1\gd_1&=&0,\nonumber\\
-h_2''+\frac{\de V_T}{\de h_2}(h_1,h_2,T) + \frac{T}{2}h_2 G_2\gd_2&=&0\label{fl4},
\end{eqnarray}
with 
\begin{equation}
G_1 =\begin{bmatrix} 
N_W c_{1-}g^2/2\\ 
N_W c_{2-}g^2/2\\
N_W c_{2-}g^2/2\\
 0\\
 0\\
 0\\
 0\\
 0\\
 0  
 \end{bmatrix},\ G_2 =\begin{bmatrix} 
N_W c_{1-}g^2/2\\ 
N_W c_{2-}g^2/2\\
N_W c_{2-}g^2/2\\
N_{\tilde{t}} c_{1-}y_t^2\\
N_{\tilde{t}} c_{2-}y_t^2\\
N_{\tilde{t}} c_{2-}y_t^2\\
N_t c_{1+} y_t^2 \\
N_t c_{2+} y_t^2 \\
N_t c_{2+} y_t^2 
\end{bmatrix},\   \gd_1=\begin{bmatrix} 
   \delta\mu_W \\ 
  \gd T_W \\ 
T\gd v_W\\
  \delta\mu_{{\tilde t}}\\ 
   \gd T_{{\tilde t}} \\ 
T\gd v_{{\tilde t}}\\
  \delta\mu_t \\  
  \gd T_t\\
 T\gd v_t
  \end{bmatrix}\!,
\ \mbox{and}\ 
  \gd_2=\begin{bmatrix} 
   \delta\mu_W \\ 
\gd T_W\\
 T\gd v_W\\
  0\\
  0\\
  0\\
  0\\
  0\\
  0
  \end{bmatrix}.\label{motion}
\end{equation}
The vectors $\delta_1$, $\delta_2$ in eq.~(\ref{motion}) contain the
perturbation functions which can be found as solution of
eq.~(\ref{fluidfinal}). Here we used the splitting of the perturbation
vector we discussed at eq.~(\ref{msquared}). Thus one is lead to a
system of linear, coupled differential equations. It can be solved
numerically. The solutions give the profiles of the perturbations in
the wall. Such profiles are discussed in \cite{Pantano} for single
scalar models. We now use again the thick wall limit. First we now
approximate $\gd'=0$. (This will be improved later on). Then we can
solve eq.~(\ref{fluidfinal}) for the perturbation vectors
$\delta_{1,2}$:
\begin{equation}
\delta_{1,2}=\Gamma^{-1}F_{1,2}.\label{solve}
\end{equation}
We obtain $F_{1,2}$ for stops, top and $W$ from eq.~(\ref{fluidfinal}) in
the simplified form
\begin{equation}
F_2=\frac{v_w}{2T}\begin{bmatrix}
c_{1-}(m_W^2)'\\
c_{2-}(m_W^2)'\\
0\\
c_{1-}(m_{\tilde{t}}^2)'\\
c_{2-}(m_{\tilde{t}}^2)'\\
0\\
c_{1+}(m_t^2)'\\
c_{2+}(m_t^2)'\\
0
\end{bmatrix}=\frac{v_w}{2T}h_2h'_2\begin{bmatrix}
c_{1-}g^2/2\\
c_{2-}g^2/2\\
0\\
c_{1-}y^2_t\\
c_{2-}y^2_t\\
0\\
c_{1+}y^2_t\\
c_{2+}y^2_t\\
0
\end{bmatrix}\equiv\frac{v_w}{2T}h_2h'_2 \tilde{F}_2,\label{motion2}
\end{equation}
and 
\begin{equation}
F_1=\frac{v_w}{2T}\begin{bmatrix}
c_{1-}(m_W^2)'\\
c_{2-}(m_W^2)'\\
0
\end{bmatrix}=\frac{v_w}{2T}h_1h'_1
\begin{bmatrix}
c_{1-}g^2/2\\
c_{2-}g^2/2\\
0
\end{bmatrix}\equiv\frac{v_w}{2T}h_1h'_1 \tilde{F}_1.\label{motion3}
\end{equation}
Finally, we arrive at the bubble wall equations of motion with friction,
\begin{eqnarray}
h_1^{\prime\prime}-V'_T&=&\eta_1v_w\frac{h_1^2}{T}h_1',\nonumber\\
h_2^{\prime\prime}-V'_T&=&\eta_2v_w\frac{h_2^2}{T}h_2'.\label{weqmo}
\end{eqnarray}
The dimensionless constants $\eta_1$ and $\eta_2$ are defined through
the relation
\begin{eqnarray}
\eta_1 &=& \frac{T}{4} G_1\Gamma^{-1}\tilde{F}_1\label{eta1},\\
\eta_2 &=& \frac{T}{4} G_2\Gamma^{-1}\tilde{F}_2\label{eta2},
\end{eqnarray}
where we used the definitions for $G_{1,2}$ in eq.~(\ref{motion}) and
$\tilde{F}_{1,2}$ from eqs.~(\ref{motion2}) and (\ref{motion3})
above. The factor $T$ rescales $\Gamma^{-1}\sim T^{-1}$.  The friction
terms $\eta_1$ and $\eta_2$ are still slightly $\tb$-dependent. The
constants give the viscosity of the medium which is perturbed by the
moving wall surface. But since $\delta'$ in general is not negligible
we need the full solution to eq.~(\ref{fluidfinal}) and $\eta_{1,2}$
cannot be defined as in eqs.~(\ref{eta1}) and (\ref{eta2}) and the
r.h.s. of (\ref{weqmo}) maintain an implicit $v_w$-dependence beside
the factor.  We then have to solve eq.~(\ref{fl4}) directly instead of
eq.~(\ref{weqmo}).  In the next section we proceed with
(\ref{weqmo})in order to obtain an analytical formula for the wall
velocity which is useful in any case.
%
\section{Wall Velocity in the MSSM}
\label{sec:wv}
In order to solve eqs.~(\ref{weqmo}) we derive a virial theorem, based
on the necessity that for a stationary wall the pressure to the wall
surface is balanced by the friction: the pressure from inside the
bubble which is responsible for the expansion and the pressure
resulting from the viscosity of the plasma must be equal.  The
pressure on a free bubble wall can be obtained from l.h.s. of the
equation of motion (\ref{weqmo}) by
\begin{equation}
p_1=\int_0^{\infty}\left(h_1^{\prime\prime}-\frac{\de V_T}{\de h_1}\right)h_1'dx=V_T(h_1(0))-V_T(h_1(x=\infty))=\Delta V_T,
\end{equation}
which is compensated due to the friction term, the r.h.s. of (\ref{weqmo}):
\begin{equation}
\int\left(h_1^{\prime\prime}-\frac{\de V_T}{\de h_1}\right)h_1'dx= \int\eta_1v_w\frac{h_1^2}{T}(h'_1)^2=\Delta V_T,
\end{equation}
where $\Delta V_T$ is the difference in the effective potential values
at the transition temperature $T_n$, which is basically the nucleation
temperature.

Since we have two equations of the same type for each of the Higgs
scalars and both of them develop friction terms, we have to add the
pressure on the bubble surface. They are different due to the
different particle species and couplings. Thus we find different
pressures which would lead to different wall velocities, if the
equations were completely decoupled. But due to the effective
potential which couples the equations of motion we have
back-reaction. This leads to a change in
$\Delta\beta=\mbox{max}(\de\beta/\de z)$.  It might be interesting to
investigate this question in more detail, since a larger $\Delta\beta$
is highly welcome to obtain a larger baryon asymmetry. This point gets
even more important with the knowledge of the results of
\cite{John4,JohnPhD} where we realized that in the MSSM transitional
CP violation does not occur. Therefore we must exploit the explicit
phases which may nevertheless be strongly restricted by experimental
bounds. The determination of $\Delta\beta$ may be done numerically by
solving eqs.~(\ref{weqmo}) with extensions of the methods of
\cite{John3,JohnPhD,JohnProc}. 

For our estimate of the wall velocity we will use as approximation a
constant $\tb$ since the deviation is so strongly suppressed as
found in \cite{SecoNum,Cline12,John3}. Then we can
add the expressions for the pressure for both Higgs fields, $p_1$ and
$p_2$:
\begin{equation}
p_1+p_2=\frac{v_w}{T}\left\{\eta_1\int h_1^2(h_1')^2dx +\eta_2\int h_2^2(h_2')^2dx\right\} = 2\Delta V_T.\label{p1p2}
\end{equation}
Next, we define an average direction and introduce a field
$h=\sqrt{h_1^2+h_2^2}$. We find
\begin{equation}
\frac{2\Delta V_T T}{v_w}=\sin^4\!\beta(\eta_2+\cot^4\!\beta\eta_1)\int h^2(h')^2dx.
\label{combi}
\end{equation}
We realize that the term corresponding to $\eta_1$, in comparison to
the second, is strongly suppressed. Already moderately small values of
$\tb$=2,(3,6) cause roughly a suppression of order 10,(100,1000).  The
wall velocity is therefore predominantly determined by the second
Higgs.  Moreover, this behaviour also is supported by the strong
Yukawa couplings of the stop and the top which couple asymmetric in
favour of $h_2$. A `large' $\tb\gsim 2$ leads to a friction term,
which is solely determined by $\eta_2$.  The limit $\tb\rightarrow 0$
leads to a large but finite velocity determined by the ($\tb$
dependent) $\eta_1$. Nevertheless, due to a different particle content
$\eta_1$ does not recover SM friction in this limit. However, the
lower experimental limit is $\tb\gsim 2$.

In \cite{SecoNum,John3} it was realized that in the MSSM the
approximation to the bubble wall profile by a kink is a rather good
choice. Hence, we set for the common Higgs field $h(x)$
\begin{equation}
h(x)=\frac{h_{crit}}{2}\left(1+\tanh\frac{x}{L_w}\right),
\end{equation}
where $L_w$ is the wall thickness and $h_{crit}$ is the nontrivial vev
at the nucleation temperature $T_n$.  The integral in
eq.~(\ref{combi}) is evaluated as
\begin{equation}
\int h^2(h')^2dx=\frac{h_{crit}^4}{10L_w},
\end{equation}
leading to the desired  equation for the wall velocity
\begin{equation}
v_w=\frac{20L_w}{h_{crit}^4}\frac{\Delta V_T(T_n) T_n}{\sin^4\!\beta(\eta_2+\eta_1\cot^4\!\beta)}.
\label{master}
\end{equation}
The missing numbers for $L_w$, $h_{crit}$, $T_n$, and $\Delta V(T_n)$
can be independently determined with methods described in
\cite{John3,JohnPhD,SecoNum,MooreServant}.
\section{Viscosity and Wall Velocity}
\label{sec:graphs}
The next step is to determine the specific number for the friction
terms. Therefore we have to calculate scattering and annihilation
rates resulting from the corresponding Feynman graphs.  First we
consider strong interactions. These processes contribute to order
$g^4$. We will drop all terms of order $m/T$ and therefore also
s-channel processes which are of order $m^2/T^2$ (cf. \cite{MP12}).
The scattering amplitudes are given by the following graphs. In
principle, a stop can scatter off quarks, squarks and gluons:
\vspace{-1cm}
\begin{eqnarray}
\SetScale{1.0}
\begin{picture}(100,85)(-40,0)
\Line(0,15)(-20,35)
\Text(-23,38)[r]{ $\tilde{t}_R$}
\Line(0,15)(20,35)
\Text(23,38)[c]{ $\tilde{t}_R$}
\Line(0,-15)(-20,-35)
\Text(-23,-38)[r]{ $\tilde{q}$}
\Line(0,-15)(20,-35)
\Text(23,-38)[c]{ $\tilde{q}$}
\Gluon(0,15)(0,-15) 3 6
\end{picture}
&
\begin{picture}(120,85)(-40,0)
\Line(0,15)(-20,35)
\Text(-23,38)[r]{ $\tilde{t}_R$}
\Line(0,15)(20,35)
\Text(23,38)[c]{ $\tilde{t}_R$}
\Line(0,-15)(-20,-35)
\Text(-23,-38)[r]{ $q$}
\Line(0,-15)(20,-35)
\Text(23,-38)[c]{ $q$}
\Gluon(0,15)(0,-15) 3 6
\end{picture}
&
\begin{picture}(100,85)(-40,0)
\Line(0,15)(-20,35)
\Text(-23,38)[r]{ $\tilde{t}_R$}
\Line(0,15)(20,35)
\Text(23,38)[c]{ $\tilde{t}_R$}
\Gluon(0,-15)(-20,-35) 3 6
\Text(-23,-38)[r]{$g$}
\Gluon(0,-15)(20,-35) 3 6
\Text(23,-38)[l]{$g$}
\Gluon(0,15)(0,-15) 3 6
\end{picture}
\label{scat0}
\\
\begin{picture}(100,85)(-40,0)
\Line(0,0)(-20,35)
\Text(-23,38)[r]{$\tilde{t}_R$}
\Line(0,0)(20,35)
\Text(23,38)[l]{$\tilde{t}_R$}
\Gluon(0,0)(-20,-35) 3 7
\Text(-23,-38)[r]{$g$}
\Gluon(0,0)(20,-35) 3 7
\Text(23,-38)[l]{$g$}
\end{picture}
&
\begin{picture}(120,85)(-40,0)
\Line(-15,0)(-35,20)
\Text(-38,23)[r]{ $\tilde{t}_R$}
\Gluon(-15,0)(-35,-20)3 6 
\Text(-38,-23)[r]{$g$}
\Line(15,0)(35,20)
\Text(38,23)[c]{ $\tilde{t}_R$}
\Gluon(15,0)(35,-20) 3 6 
\Text(38,-23)[c]{ $g$}
\Line(15,0)(-15,0)
\end{picture}
&
\begin{picture}(120,85)(-40,0)
\Line(-20,0)(35,20)
\Text(-38,23)[r]{ $\tilde{t}_R$}
\Gluon(-35,-20)(-20,0) 3 6
\Text(-38,-23)[r]{ $g$}
\Line(20,0)(-35,20)
\Text(38,23)[c]{ $\tilde{t}_R$}
\Gluon(35,-20)(20,0) 3 6 
\Text(38,-23)[c]{ $g$}
\Line(20,0)(-20,0)
\end{picture}
\label{scat1}\\
&& \nonumber\\
&& \nonumber
\end{eqnarray}
The annihilation graphs can be obtained by exchanging the s and t
direction.  In our approximation, there are only a few contributions
to the matrix elements.  For instance, we assume all the squarks
besides the stop to be superheavy and decoupled. Hence, the first
graph in (\ref{scat0}) would only contribute for the stop-stop
scattering. But, as the stops are at the same temperature and
velocity, there is no change in the perturbations:
$\sum\delta=0\mu+(E_p-E_{p'})\delta T +(p_z-p_z')\delta v=0$.
Therefore, stop-squark scattering does not contribute to the rates in
this approximation.

The corresponding types of graphs have to be taken into account for
the rates of the top-quark. They can scatter off gluons, quarks and
squarks. Compared to the SM, a new scattering graph appears for the
top which in the MSSM may scatter off squarks.

We calculate the rates which are defined in eq.~(\ref{gammadef}) by
the integration of the collision integral $C[f]$, hence we need
\begin{eqnarray}
C[f] &=& \int\frac{d^3kd^3p'd^3k'}{(2\pi)^92E_p2E_{p'}2E_{k'}}\left|{\cal M}\right|^2\times\nonumber\\
&& \times (2\pi)^4\delta(p+k-p'-k'){\cal P}[f]\nonumber\\
\!\!\!\!\mbox{with} \qquad{\cal P}[f] &=&f_pf_k(1\pm f_{p'})(1\pm f_{k'})-f_{p'}f_{k'}(1\pm f_{p})(1\pm f_{k}).
\end{eqnarray}
The sign depends on the statistics of incoming an outgoing particles.
For stop gluon interactions we have a minus sign in any case. The
integrals are calculated in the relativistic limit. In this
approximation we find the following matrix elements (after averaging
over outgoing particles). For annihilation of stops into gluons/ghosts we have
\begin{equation}
|{\cal M}_A|^2=\frac{137}{9}g_s^4+36g_s^4\frac{st+t^2}{s^2},
\end{equation}
and for stops scattering off gluons/ghosts, 
\begin{equation}
|{{\cal M}}_S|^2=\frac{128}{9}g_s^4+32g_s^4\frac{st+s^2}{t^2}
\end{equation} 
which can be understood easily by crossing the annihilation result and
taking into account that stops scatter off ghosts and anti-ghosts. The
large numbers result from large color and momentum factors each.  We
include thermal masses of the exchange particle $m_p$ in the
denominator to regularize the divergence for small $t$. Different from
\cite{MP12} we we fully integrate the rates numerically.  Some details
of the integration of the rates are given in App. A. For the numerical
evaluation of the resulting integrals we use
$g_s^2/(4\pi)=\alpha_s=0.12$, $\alpha_W=1/30$. The integrals for
top-W-scattering can be found in \cite{MP12}. Nevertheless, we have to
reevaluate those expressions with different plasma masses for tops, W
bosons, and stops since the supersymmetric plasma differs from the SM
plasma \cite{erv}. We use as plasma mass of the stop for our
scenario\cite{Mikko3dCP}
\begin{equation}
m^2_{\tilde{t}_R}=-m_U^2+(\frac{2}{3}g_s^2+\frac{1}{3}h_t^2-\frac{1}{6}(A_t^2+\mu^2)/m_Q^2)T^2.
\end{equation}
As a required condition for baryogenesis we need very negative
$-m_U^2$. As plasma mass for the gluon we take $m^2_{g}\approx
2/3g_s^4T^2$. We use $m_W^2=5/6g^2T^2$ for the $W$-mass,
$m_t^2=(1/6g_s^2+3/32g^2+1/8h_t^2)T^2$ for the top
mass~\cite{erv}. Annihilations into leptons and quarks include lepton
masses $m_l=3/32g^2T^2$ and bosons with mass $m_q^2=1/6g^2T^2$ leading
to an effective plasma mass $\ln<\!m^2\!>=3/4\ln m_q^2+1/4\ln m_l^2$.

After eliminating the perturbations with help of the rates as shown
above in eqs.~(\ref{fluidfinal}), (\ref{solve}) and finally
eqs.~(\ref{eta1}) and (\ref{eta2}), we can determine $\eta_1$ and
$\eta_2$.  Numerically we find for $\tb\gsim {\cal O}(2)$ and $m_U\lsim{\cal
O}(0~\text{GeV})$
\begin{eqnarray}
\eta_1 &\approx& {\cal O}(0.1),\\
\eta_2 &\leq& {\cal O}(100).
\end{eqnarray}
This emphasizes the domination of $\eta_2$.  In the Standard Model
with only one viscosity constant one finds roughly $\eta_{SM}\approx
3$.  

The wall velocities are to be determined from the value of $\tb$ at
the transition temperature $T_n$ and the corresponding values of
$L_w$, $h_{crit}$ and the difference in the potential heights $\Delta
V_T(T_n)$ from the one-loop resummed potential.
\begin{figure}[tb]

\vspace*{.5cm}

\hspace*{2cm}
\epsfysize=7cm\epsffile{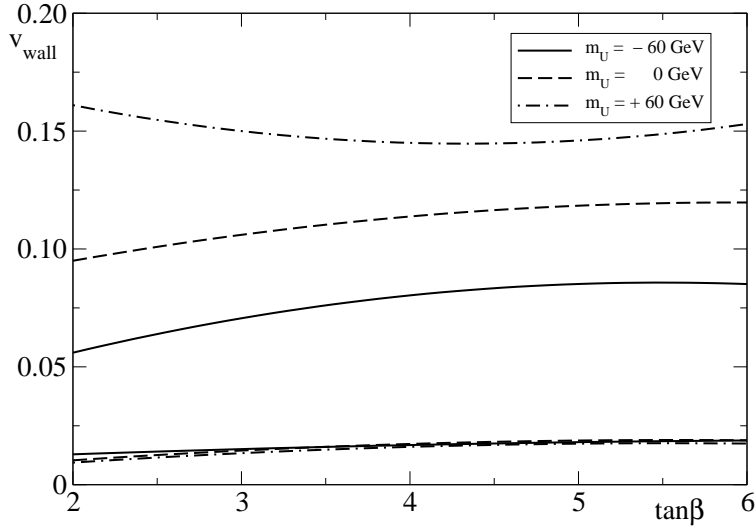}

\caption{Wall velocity in dependence on the parameter $\tb(T\!=\!0)$
for $m_Q=2$TeV, $A_t=\mu=0$, and $m_A=400$GeV for $m_U^2=-60^2, 0,
60^2$GeV${}^2$. Lower bunch of graphs for $\gd'=0$, upper for
$\gd'\neq 0$.}
\label{fig:wallbeta}
\end{figure}
In the whole scenario we assume the left handed stops to be decoupled
and the right handed ones very light. We have no mixing $A_t=\mu=0$,
$m_Q=2TeV$, and $m_A=400GeV$. For a strong PT we prefer small right
handed stop parameters $m_U^2$.  The value $m^2_U=-(60GeV)^2$, thus
$m_{\tilde{t}}=161 GeV$, is very near to the triple
point\cite{John2,MooreServant}. There we find a strong phase
transition with $v/T=0.96$ at one-loop level for $\tb=2.0$. At
two-loop level it is even larger permitting larger $\tb$ for the same
strength~\cite{John2,litestop}. In this parameter range, eventually
including some stop mixing, one can reach Higgs masses up to
$m_H\approx 110$~GeV which agree with the present experimental bound.

\begin{figure}[tb]
\vspace*{.5cm}

\hspace*{2cm}
\epsfysize=7cm{\epsffile{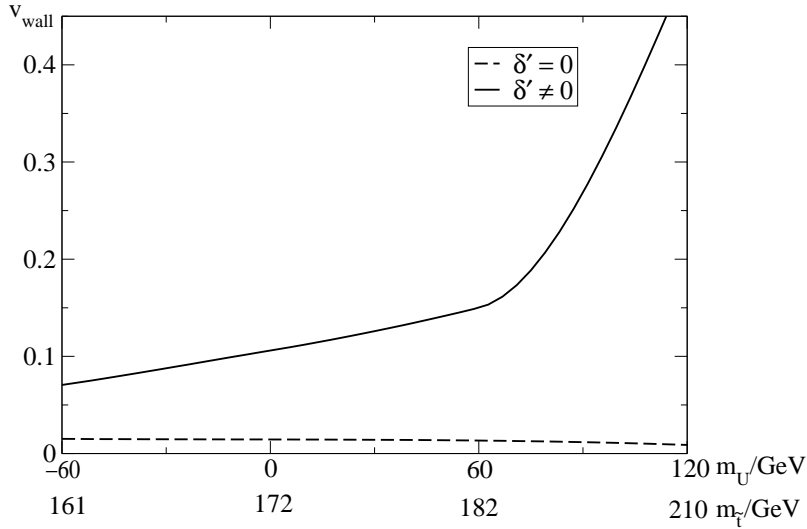}}

\caption{Wall velocity for increasing stop mass parameter $m_U$. At
very large $m_U$ (physically disfavoured) the stops decouple from the
plasma leading to a large velocity while the first approximation
 $\gd'=0$ contrarily leads to decreasing $v_W$. The diagram is
calculated for $\tb=3$, $A_t=\mu=0$, and $m_A=400$GeV.}
\label{fig:mU}
\end{figure}

Performing all the calculations the wall velocity turns out to be of
the order of $v_w\approx 0.05-0.1$ which is lower than in the Standard
Model. Due to the large numerical factor of the matrix elements, the
overall rates are larger than expected but probably sufficiently small
for baryogenesis.
In Fig.~\ref{fig:wallbeta} the dependence on the parameter $\tb(T=0)$
is shown in the physically interesting range $2\leq \tb \leq 6$ for
two values of $m_U^2=-60^2,0,60^2$ GeV${}^2$. The lower bunch of
curves is determined for our first approximation $\delta'=0$ the upper
from the full numerical solution to eq.~(\ref{fluidfinal}) with
$\delta'\neq 0$. We find\footnote{Since eq.~(\ref{fluidfinal}) depends
through ${\mathbb{A}}$ on $v_w$ we solved it iteratively.}
corrections of roughly a factor of three for small $m_U$ (light stop).
Very heavy stops should decouple more and more which would lead to to
increasing wall velocities again. This behaviour can reproduced with
the full solution.  In Fig.~\ref{fig:mU} this principle behaviour is
demonstrated over a wide range of $m_U$ and $m_{\tilde{t}_R}$ for
$\tb=3$.
Nevertheless, for large (positive) $m_U^2$ the used approximations
become worse and corrections of order ${\cal{O}}(m^2/T^2)$ are
important. Then our approximations break down. However, this range is
in any case physically disfavoured for a strong PT.
\section{Conclusion and Discussion}
The velocity of walls of expanding bubbles during the phase transition
is an important ingredient for the calculation of the baryon asymmetry
at the electroweak scale. In this paper we developed the utilities for
the calculation of the wall velocity in extensions of the Standard
Model and applied them to the MSSM. We generalized the method of
\cite{MP12} to include more than one Higgs field and arbitrarily many
particles and applied it to the MSSM with a light right handed stop.

We estimated the effect of the stops on the wall velocity in the
plasma of the simplest supersymmetric extension of the SM with all
particles heavy or in equilibrium besides the light stop, the $W$
bosons and the top quark. With our calculation we give estimates on
the friction or viscosity coefficients of this plasma. There are two
parts, one for the two CP even Higgs fields each.  In the region of
$\tb$ which is interesting for baryogenesis the velocity depends
moderately on $\tb$ and we find values of $v_w\approx (5-10)\cdot
10^{-2}$.  We can answer the question of the title with {\em Yes}:
stops {\em do} slow down the bubble wall.

We have not included effects of further out-of-equilibrium SUSY
particles like charginos and neutralinos and gave only the leading
order results. Gluinos may have considerable influence due to their
multiple interactions. We estimated the possible effect of a light
Higgsino which produce logarithmically dominating equilibration rates.
There may be a strong effect, reducing the bubble wall velocity.

Also a more precise consideration of the mass of the outer legs in the
graphs presumably give an effect. We calculated the matrix elements in
the massless limit, but especially for the stop this might be
questionable. Moreover we neglected possible effects on the transition
parameters due to a space dependent background temperature. A change
in the nucleation temperature $T_n$ and potentially large reheating
might change the results.  These effects may also affect a deviation
of the bubble wall profile from the kink Ansatz, as discussed
in~\cite{MP12}.

The contribution of the stop in our WKB calculation appears to be at
least of the same order of magnitude as the soft gauge boson
contribution estimated in \cite{Moorewall2000}. A further calculation
using combined techniques including B\"odeker's effective theory
\cite{bodeker} were nevertheless useful.

Low wall velocities agree with the requirements of baryogenesis and
enlarge the baryon asymmetry.  Our results agree with the possibility
that electroweak baryogenesis is a realistic scenario for baryon
asymmetry of our Universe.

\paragraph*{Acknowledgements.}
We thank D.~B\"odeker, A. Hebecker, S.~J.~Huber, B.~Kastening,
M.~Laine, and S.~Weinstock for useful discussions and G.~Moore for his
constructive criticism on an earlier version of this paper. This work
was partly supported by the TMR network {\em Finite Temperature Phase
Transitions in Particle Physics}, EU contract no.\ FMRX-CT97-0122.
\appendix
\renewcommand{\thesection}{Appendix~\Alph{section}}
\renewcommand{\theequation}{A.\arabic{equation}}
\section{Strong interactions}
\label{app:strong}
The matrix elements for stop-gluon interactions after averaging over
outgoing particles are
\begin{equation}
|{{\cal M}}_A|^2=\frac{137}{9}g_s^4+36g_s^4\frac{st+s^2}{t^2}, \quad\mbox{and} \quad
|{{\cal M}}_S|^2=\frac{128}{9}g_s^4+32g_s^4\frac{st+t^2}{s^2}
\end{equation} 
for annihilation and scattering, respectively.  These matrix elements
have to be integrated for the rates.  We use the formalism and the
definitions of \cite{Moore2001}, App. A, for both, annihilation
(s-channel) and scattering (t-channel) rates.
\paragraph{Annihilation} 
We rewrite $s$, $t$, $u$ with the following identities
\begin{eqnarray}
&& s=\omega^2-q^2,\quad u=-s-t,\quad k=\omega-p,\quad k'=\omega-p',\\
&& t=\frac{s}{2q^2}\left\{\left[(p-k)(p'-k')-q^2\right]+\cos\phi\sqrt{(4pk-s)(4p'k'-s)}\right\}.
\end{eqnarray}
Then the rates for annihilation are given by
\begin{eqnarray}
\Gamma_{\mu 1} & = & \frac{1}{T^3}\int dR 2 |{{\cal M}}_A|^2(p',p,q,\omega,\phi),\\
\Gamma_{T1} & = & \frac{1}{T^4}\int dR (E_p+E_k) |{{\cal M}}_A|^2(p',p,q,\omega,\phi),\\
\Gamma_{\mu 2} & = & \frac{1}{T^4}\int dR 2E_p |{{\cal M}}_A|^2(p',p,q,\omega,\phi),\\
\Gamma_{T2} & = & \frac{1}{T^5}\int dR E_p(E_p+E_k) |{{\cal M}}_A|^2(p',p,q,\omega,\phi),\\
\Gamma_{v} & = & \frac{1}{T^5}\int dR p_z(p_z+k_z) |{{\cal M}}_A|^2(p',p,q,\omega,\phi),
\end{eqnarray}
where 
\begin{equation}
\int dR = \frac{1}{2^9\pi^6}\int_0^{2\pi}d\phi\int_0^\infty d\omega\int_0^\omega dq\int_{\frac{\omega+q}{2}}^{\frac{\omega-q}{2}}dp\int_{\frac{\omega+q}{2}}^{\frac{\omega-q}{2}}dp' f_pf_k(1+f_{p'})(1+f_{k'}).
\end{equation}
\paragraph{Scattering}
Here we rewrite $s,t,u$ with
\begin{eqnarray}
&&t=\omega^2-q^2u=-t-s,\quad p'=p+\omega,\quad k'=k-\omega,\\
&&s=\frac{-t}{2q^2}\left\{\left[(p+p')(k+k')+q^2\right]-\cos\phi\sqrt{(4pp'+t)(4kk'+t)}\right\}.
\end{eqnarray}
Then the rates for scattering are given by 
\begin{eqnarray}
\Gamma_{T2} & = & \frac{1}{T^5}\int dR E_p(E_p-E_{p'}) |{\cal M}_S|^2(p',p,q,\omega,\phi),\\
\Gamma_{v} & = & \frac{1}{T^5}\int dR p_z(p_z-p'_z) |{\cal M}_S|^2(p',p,q,\omega,\phi),
\end{eqnarray}
where 
\begin{equation}
\int dR = \frac{1}{2^9\pi^6}\int_0^{2\pi}d\phi\int_0^\infty dq\int_{-q}^{q} d\omega\int_{\frac{\omega-q}{2}}^\infty dp\int_{\frac{\omega+q}{2}}^\infty dk f_pf_k(1+f_{p'})(1+f_{k'}).
\end{equation}
The $dR$ integration must be done numerically and leads to
friction coefficients $\eta_1 \approx {\cal O}(0.1)$ and
$\eta_2\leq{\cal O}(100)$, slightly depending on the physical parameters of
the system.
\begingroup\raggedright

\end{document}